\documentclass[onecolumn]{article}  
\usepackage{epsfig,dsfont,amssymb,amsmath,amsthm,amsfonts,amsbsy,mathrsfs}

\begin{document}

\title{Tighter monogamy relations in multipartite systems}
\author{Zhi-Xiang Jin 
\thanks{Corresponding author: jzxjinzhixiang@126.com}\\
School of Mathematical Sciences,  Capital Normal University, \\ Beijing 100048,  China\\
\and Jun Li
\thanks{Corresponding author: lijunnl123@163.com}\\
School of Mathematical Sciences,  Capital Normal University,  \\Beijing 100048,  China\\
\and Tao Li
\thanks{Corresponding author: litao@btbu.edu.cn}\\
School of Science, Beijing Technology and Business University, \\Beijing 100048, China\\
\and Shao-Ming Fei
\thanks{Corresponding author: feishm@mail.cnu.edu.cn}\\
School of Mathematical Sciences,  Capital Normal University,  \\Beijing 100048,  China\\
Max-Planck-Institute for Mathematics in the Sciences, \\Leipzig 04103, Germany
}

\maketitle
\begin{abstract}
Monogamy relations characterize the distributions of entanglement in multipartite systems. We investigate monogamy relations related to the concurrence $C$, the entanglement of formation $E$, negativity $N_c$ and Tsallis-$q$ entanglement $T_q$. Monogamy relations for the $\alpha$th power of entanglement have been derived, which are tighter than the existing entanglement monogamy relations for some classes of quantum states. Detailed examples are presented.
\end{abstract}

\section{INTRODUCTION}
Due to the essential roles played in quantum communication and quantum information processing, quantum entanglement \cite{1,2,3,4,5,6,7,8} has been the subject of many recent studies in recent years. The study of quantum entanglement from various viewpoints has been a very active area and has led to many impressive results. As one of the fundamental differences be tween quantumand classical correlations, an essential property of entanglement is that a quantum system entangled with one of the other subsystems limits its entanglement with the remaining ones. The monogamy relations give rise to the distribution of entanglement in the multipartite quantum systems.Moreover, themonogamy property has emerged as the ingredient in the security analysis of quantum key distribution \cite{9}.	

For a tripartite system $A$, $B$, and $C$, the usual monogamy of an entanglement measure $\mathcal{E}$ implies that \cite{10} the entanglement between $A$ and $BC$ satisfies $\mathcal{E}_{A|BC}\geqslant \mathcal{E}_{AB} +\mathcal{E}_{AC}$. However, such monogamy relations are not always satisfied by all entanglement measures for all quantum states. In fact, it has been shown that the squared concurrence $C^2$ \cite{11,12} and entanglement of formation $E^2$ \cite{13} satisfy the monogamy relations for multiqubit states. The monogamy inequality was further generalized to various entanglement measures such as continuous-variable entanglement \cite{14,15,16}, squashed entanglement \cite{10,17,18}, entanglement negativity \cite{19,20,21,22,23}, Tsallis-$q$ entanglement \cite{24,25}, and Renyi entanglement \cite{26,27,28}.

In this paper, we derive monogamy inequalities which are tighter than all the existing ones, in terms of the concurrence $C$, the entanglement of formation $E$, negativity $N_c$, and Tsallis-$q$ entanglement $T_q$.

\section{TIGHTER MONOGAMY RELATIONS FOR CONCURRENCE}

We first consider the monogamy inequalities satisfied by the concurrence. Let $\mathds{H}_X$ denote a discrete finite-dimensional complex vector space associated with a quantum subsystem $X$. For a bipartite pure state $|\psi\rangle_{AB}$ in vector space $\mathds{H}_A\otimes \mathds{H}_B$, the concurrence is given by \cite{29,30,31}
$
\label{CD} C(|\psi\rangle_{AB})=\sqrt{{2\left[1-\mathrm{Tr}(\rho_A^2)\right]}},
$
where $\rho_A$ is the reduced density matrix by tracing over the subsystem $B$, $\rho_A=\mathrm{Tr}_B(|\psi\rangle_{AB}\langle\psi|)$. The concurrence for a bipartite mixed state $\rho_{AB}$ is defined by the convex roof extension
$
 C(\rho_{AB})=\min_{\{p_i,|\psi_i\rangle\}}\sum_ip_iC(|\psi_i\rangle),
$
where the minimum is taken over all possible decompositions of $\rho_{AB}=\sum\limits_{i}p_i|\psi_i\rangle\langle\psi_i|$, with $p_i\geqslant0$, $\sum\limits_{i}p_i=1$ and $|\psi_i\rangle\in \mathds{H}_A\otimes \mathds{H}_B$.

For a tripartite state $|\psi\rangle_{ABC}$,the concurrence of assistance is defined by \cite{32,33}
\begin{eqnarray*}
C_a(|\psi\rangle_{ABC})\equiv C_a(\rho_{AB})=\max\limits_{\{p_i,|\psi_i\rangle\}}\sum_ip_iC(|\psi_i\rangle),
\end{eqnarray*}
where the maximum is taken over all possible decompositions of $\rho_{AB}=\\ \mathrm{Tr}_C(|\psi\rangle_{ABC}\langle\psi|)=\sum\limits_{i}p_i|\psi_i\rangle_{AB}\langle\psi_i|.$ When $\rho_{AB}=|\psi\rangle_{AB}\langle\psi|$ is a pure state, one has $C(|\psi\rangle_{AB})=C_a(\rho_{AB})$.

For an $N$-qubit state $\rho_{AB_1\cdots B_{N-1}}\in \mathds{H}_A\otimes \mathds{H}_{B_1}\otimes\cdots\otimes \mathds{H}_{B_{N-1}}$, the concurrence $C(\rho_{A|B_1\cdots B_{N-1}})$ of the state $|\psi\rangle_{A|B_1\cdots B_{N-1}}$, viewed as a bipartite state under the partition $A$ and $B_1,B_2,\cdots, B_{N-1}$, satisfies \cite{34}
\begin{eqnarray}\label{mo1}
&&C^{\alpha}(\rho_{A|B_1,B_2\cdots,B_{N-1}})\nonumber \\
&&\geqslant C^{\alpha}(\rho_{AB_1})+C^{\alpha}(\rho_{AB_2})+\cdots+C^{\alpha}(\rho_{AB_{N-1}}),
\end{eqnarray}
for $\alpha\geqslant2$, where $\rho_{AB_i}=\mathrm{Tr}_{B_1\cdots B_{i-1}B_{i+1}\cdots B_{N-1}}(\rho_{AB_1\cdots B_{N-1}})$. The relation ($\ref{mo1}$) is further improved so that for $\alpha\geqslant2$, if $C(\rho_{AB_{i}})\geqslant C(\rho_{A|B_{i+1}\cdots B_{N-1}})$ for $i= 1,2, \cdots, m$ and $C(\rho_{AB_{j}})\leqslant C(\rho_{A|B_{j+1}\cdots B_{N-1}}$ for $j= m+1, \cdots, N-2$, $\forall 1 \leqslant m\leqslant N-3$, $N\geqslant 4$, then \cite{35},
\begin{eqnarray}\label{mo2}
&&C^\alpha(\rho_{A|B_1B_2\cdots B_{N-1}})\geqslant \nonumber\\
&&C^\alpha(\rho_{AB_1})+\frac{\alpha}{2} C^\alpha(\rho_{AB_2})+\cdots+\left(\frac{\alpha}{2}\right)^{m-1}C^\alpha(\rho_{AB_m})\nonumber\\
&&+\left(\frac{\alpha}{2}\right)^{m+1}\left[C^\alpha(\rho_{AB_{m+1}})+\cdots+C^\alpha(\rho_{AB_{N-2}})\right]\nonumber\\
&&+\left(\frac{\alpha}{2}\right)^{m}C^\alpha(\rho_{AB_{N-1}})
\end{eqnarray}
and for all $\alpha<0$,
\begin{eqnarray}\label{l2}
\setlength{\belowdisplayskip}{3pt}
&&C^\alpha(\rho_{A|B_1B_2\cdots B_{N-1}})< \nonumber \\
&&K[C^\alpha(\rho_{AB_1})+C^\alpha(\rho_{AB_2})+\cdots+C^\alpha(\rho_{AB_{N-1}})],
\end{eqnarray}
where $K=\frac{1}{N-1}$.

In the following, we show that these monogamy inequalities satisfied by the concurrence can be further refined and become even tighter. For convenience, we denote $C_{AB_i}=C(\rho_{AB_i})$ the
concurrence of $\rho_{AB_i}$ and $C_{A|B_1,B_2,\cdots,B_{N-1}}=C(\rho_{A|B_1 \cdots B_{N-1}})$. We first introduce two lemmas.

\textit{Lemma 1}. For any real number $x$ and $t$, $0\leqslant t \leqslant 1$, $x\in [1, \infty]$, we have $(1+t)^x\geqslant 1+(2^{x}-1)t^x$.

\textit{Proof}. Let $f(x,y)=(1+y)^x-y^x$ with $x\geqslant 1,~y\geqslant 1$, then, $\frac{\partial f}{\partial y}=x[(1+y)^{x-1}-y^{x-1}]\geqslant 0$. Therefore, $f(x,y)$ is an increasing function of $y$, i.e., $f(x,y)\geqslant f(x,1)=2^x-1$. Set $y=\frac{1}{t},~0<t\leqslant 1$, and we obtain $(1+t)^x\geqslant 1+(2^x-1)t^x$. When $t=0$, the inequality is trivial.~~~~~~~~~~~~ $\blacksquare$

\textit{Lemma 2}. For any $2\otimes2\otimes2^{n-2}$ mixed state $\rho\in \mathds{H}_A\otimes \mathds{H}_{B}\otimes \mathds{H}_{C}$, if $C_{AB}\geqslant C_{AC}$, we have
\begin{equation}\label{le2}
  C^\alpha_{A|BC}\geqslant  C^\alpha_{AB}+(2^{\frac{\alpha}{2}}-1)C^\alpha_{AC},
\end{equation}
for all $\alpha\geqslant2$.

\textit{Proof}. It has been shown that $C^2_{A|BC}\geqslant C^2_{AB}+C^2_{AC}$ for arbitrary $2\otimes2\otimes2^{n-2}$ tripartite state $\rho_{ABC}$ \cite{11,37}. Then, if $C_{AB}\geqslant C_{AC}$, we have
\begin{eqnarray*}
  C^\alpha_{A|BC}&&\geqslant (C^2_{AB}+C^2_{AC})^{\frac{\alpha}{2}}\\
  &&=C^\alpha_{AB}\left(1+\frac{C^2_{AC}}{C^2_{AB}}\right)^{\frac{\alpha}{2}} \\
  && \geqslant C^\alpha_{AB}\left[1+(2^{\frac{\alpha}{2}}-1)\left(\frac{C^2_{AC}}{C^2_{AB}}\right)^{\frac{\alpha}{2}}\right]\\
  &&=C^\alpha_{AB}+(2^{\frac{\alpha}{2}}-1)C^\alpha_{AC}
\end{eqnarray*}
where the second inequality is due to Lemma 1. As the subsystems $A$ and $B$ are equivalent in this case, we have assumed that $C_{AB}\geqslant C_{AC}$ without loss of generality. Moreover,
if $C_{AB}=0$ we have $C_{AB}=C_{AC}=0$. That is to say the lower bound becomes trivially zero.~~~~~~~~~~~~$\blacksquare$

From Lemma 2, we have the following theorem.

\textit{Theorem 1}. For an $N$-qubit mixed state, if ${C_{AB_i}}\geqslant {C_{A|B_{i+1}\cdots B_{N-1}}}$ for $i=1, 2, \cdots, m$, and
${C_{AB_j}}\leq {C_{A|B_{j+1}\cdots B_{N-1}}}$ for $j=m+1,\cdots,N-2$,
$\forall$ $1\leq m\leq N-3$, $N\geqslant 4$, then we have
\begin{eqnarray}\label{th1}
&&C^\alpha_{A|B_1B_2\cdots B_{N-1}} \nonumber \\
&&~~~\geqslant  C^\alpha_{AB_1}+(2^{\frac{\alpha}{2}}-1) C^\alpha_{AB_2}+\cdots+(2^{\frac{\alpha}{2}}-1)^{m-1}C^\alpha_{AB_m}\nonumber\\
&&~~~~~~+(2^{\frac{\alpha}{2}}-1)^{m+1}(C^\alpha_{AB_{m+1}}+\cdots+C^\alpha_{AB_{N-2}}) \nonumber\\
&&~~~~~~+(2^{\frac{\alpha}{2}}-1)^{m}C^\alpha_{AB_{N-1}}
\end{eqnarray}
for all $\alpha\geqslant2$.

\textit{Proof}. From the inequality (\ref{le2}), we have
\small
\begin{eqnarray}\label{pf11}
&&C^{\alpha}_{A|B_1B_2\cdots B_{N-1}}\nonumber\\
&&\geqslant  C^{\alpha}_{AB_1}+(2^{\frac{\alpha}{2}}-1)C^{\alpha}_{A|B_2\cdots B_{N-1}}\nonumber\\
&&\geqslant C^{\alpha}_{AB_1}+(2^{\frac{\alpha}{2}}-1)C^{\alpha}_{AB_2}
 +(2^{\frac{\alpha}{2}}-1)^2C^{\alpha}_{A|B_3\cdots B_{N-1}}\nonumber\\
&& \geqslant \cdots\nonumber\\
&&\geqslant C^{\alpha}_{AB_1}+(2^{\frac{\alpha}{2}}-1)C^{\alpha}_{AB_2}+\cdots+(2^{\frac{\alpha}{2}}-1)^{m-1}C^{\alpha}_{AB_m}\nonumber\\
&&~~~~+(2^{\frac{\alpha}{2}}-1)^m C^{\alpha}_{A|B_{m+1}\cdots B_{N-1}}.
\end{eqnarray}
\normalsize
Similarly, as ${C_{AB_j}}\leqslant {C_{A|B_{j+1}\cdots B_{N-1}}}$ for $j=m+1,\cdots,N-2$, we get
\begin{eqnarray}\label{pf12}
&&C^{\alpha}_{A|B_{m+1}\cdots B_{N-1}} \nonumber\\
&&~~~\geqslant (2^{\frac{\alpha}{2}}-1)C^{\alpha}_{AB_{m+1}}+C^{\alpha}_{A|B_{m+2}\cdots B_{N-1}}\nonumber\\
&&~~~\geqslant (2^{\frac{\alpha}{2}}-1)(C^{\alpha}_{AB_{m+1}}+\cdots+C^{\alpha}_{AB_{N-2}})\nonumber\\
&&~~~~~~~+C^{\alpha}_{AB_{N-1}}.
\end{eqnarray}
Combining (\ref{pf11}) and (\ref{pf12}), we have Theorem 1. ~~~~~~~~~~$\blacksquare$

As for $\alpha\geqslant 2$, $(2^{\frac{\alpha}{2}}-1)^m\geqslant (\alpha/2)^m$ for all $1\leq m\leq N-3$, our formula (\ref{th1}) in Theorem 1 gives a tighter monogamy relation with larger lower bounds than (\ref{mo1}), (\ref{mo2}). In Theorem 1, we have assumed that
some ${C_{AB_i}}\geqslant {C_{A|B_{i+1}\cdots B_{N-1}}}$ and some
${C_{AB_j}}\leq {C_{A|B_{j+1}\cdots B_{N-1}}}$ for the $2\otimes2\otimes\cdots\otimes2$ mixed state $\rho\in \mathds{H}_A\otimes \mathds{H}_{B_1}\otimes\cdots\otimes \mathds{H}_{{B_{N-1}}}$. If all ${C_{AB_i}}\geqslant {C_{A|B_{i+1}\cdots B_{N-1}}}$ for $i=1, 2, \cdots, N-2$, then we have the following conclusion:

\textit{Theorem 2}. If ${C_{AB_i}}\geqslant {C_{A|B_{i+1}\cdots B_{N-1}}}$ for all $i=1, 2, \cdots, N-2$, then we have
\small
\begin{eqnarray}\label{Co}
&&C^\alpha_{A|B_1\cdots B_{N-1}} \nonumber\\
&&~~~~~~\geqslant C^\alpha_{AB_1}+(2^{\frac{\alpha}{2}}-1) C^\alpha_{AB_2}+\cdots+(2^{\frac{\alpha}{2}}-1)^{N-3}C^\alpha_{AB_{N-2}}\nonumber\\
&&~~~~~~~~~+(2^{\frac{\alpha}{2}}-1)^{N-2}C^\alpha_{AB_{N-1}}.
\end{eqnarray}
\normalsize

{\it Example 1}. Let us consider the three-qubit state $|\psi\rangle$ in the generalized Schmidt decomposition form \cite{38,39},
\begin{eqnarray}\label{ex1}
|\psi\rangle&=&\lambda_0|000\rangle+\lambda_1e^{i{\varphi}}|100\rangle+\lambda_2|101\rangle \nonumber\\
&&+\lambda_3|110\rangle+\lambda_4|111\rangle,
\end{eqnarray}
where $\lambda_i\geqslant0,~i=0,1,2,3,4$ and $\sum\limits_{i=0}\limits^4\lambda_i^2=1.$
From the definition of concurrence, we have $C_{A|BC}=2\lambda_0\sqrt{{\lambda_2^2+\lambda_3^2+\lambda_4^2}}$, $C_{AB}=2\lambda_0\lambda_2$, and $C_{AC}=2\lambda_0\lambda_3$. Set $\lambda_{0}=\lambda_{1}=\frac{1}{2}$, $\lambda_{2}=\lambda_{3}=\lambda_{4}=\frac{\sqrt{6}}{6}$, one has $C_{A|BC}=\frac{\sqrt{2}}{2}$, $C_{AB}=C_{AC}=\frac{\sqrt{6}}{6}$, then $C_{A|BC}^{\alpha}=(\frac{\sqrt{2}}{2})^{\alpha}$, $C_{AB}^{\alpha}+C_{AC}^{\alpha}=2(\frac{\sqrt{6}}{6})^{\alpha}$, $C_{AB}^{\alpha}+\frac{\alpha}{2}C_{AC}^{\alpha}=(1+\frac{\alpha}{2})(\frac{\sqrt{6}}{6})^{\alpha}$, $C_{AB}^{\alpha}+(2^{\frac{\alpha}{2}}-1)C_{AC}^{\alpha}=2^{\frac{\alpha}{2}}(\frac{\sqrt{6}}{6})^{\alpha}$.
One can see that our result is better
than the results in \cite{34} and \cite{35} for $\alpha\geqslant2$; see Fig 1.
\begin{figure}
  \centering
  \includegraphics[width=7cm]{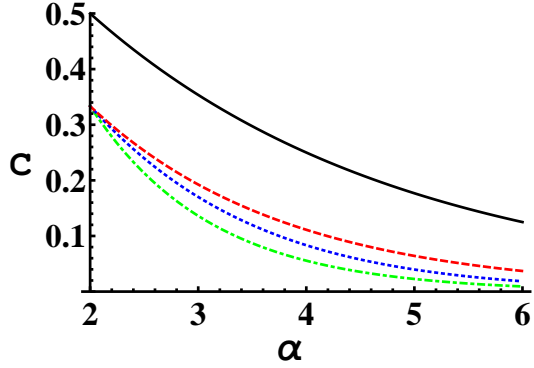}\\
  \caption{Behavior of the concurrence of $|\psi\rangle$ and its lower bound, which are functions of $\alpha$ plotted. The black solid line represents the concurrence of $|\psi\rangle$ in Example 1, the red dashed line represents the lower bound from our result, and the blue dotted (green dot-dashed) line represents the lower bound from the result in  \cite{35} (\cite{34}).}
\end{figure}

\section{TIGHTER MONOGAMY REALATIONS FOR EoF}

The entanglement of formation (EOF) \cite{40,41} is a well-defined important measure of entanglement for bipartite systems. Let $\mathds{H}_A$ and $\mathds{H}_B$ be $m$- and $n$- dimensional $(m\leqslant n)$ vector spaces, respectively. The EOF of a pure state $|\psi\rangle\in \mathds{H}_A\otimes \mathds{H}_B$ is defined by
\begin{equation}\label{SEA}
 E(|\psi\rangle)=S(\rho_A),
\end{equation}
where $\rho_A=\mathrm{Tr}_B(|\psi\rangle\langle\psi|)$ and $S(\rho)=-\mathrm{Tr}(\rho \log_2\rho).$ For a bipartite mixed state $\rho_{AB}\in \mathds{H}_A\otimes \mathds{H}_B$, the entanglement of formation is given by
\begin{equation}\label{SEB}
 E(\rho_{AB})=\min_{\{p_i,|\psi_i\rangle\}}\sum\limits_ip_iE(|\psi_i\rangle)
\end{equation}
with the minimum taking over all possible pure-state decompositions of $\rho_{AB}$.

Denote $f(x)=H\left(\frac{1+\sqrt{1-x}}{2}\right)$, where $H(x)=-x\log_2(x)-(1-x)\log_2(1-x)$. From (\ref{SEA}) and (\ref{SEB}), one has $E(|\psi\rangle)=f\left(C^2(|\psi\rangle)\right)$ for $2\otimes m~(m\geqslant2)$ pure state $|\psi\rangle$, and  $E(\rho)=f\left(C^2(\rho)\right)$ for two-qubit mixed state $\rho$ \cite{42}. It is obvious that $f(x)$ is a monotonically increasing function for $0\leqslant x\leqslant1$. $f(x)$ satisfies the following relations:
\begin{equation}\label{F2}
  f^{\sqrt{2}}(x^2+y^2)\geqslant f^{\sqrt{2}}(x^2)+f^{\sqrt{2}}(y^2),
\end{equation}
where $f^{\sqrt{2}}(x^2+y^2)=[f(x^2+y^2)]^{\sqrt{2}}.$

It has been shown that the EOF does not satisfy the inequality $E_{AB}+E_{AC}\leq E_{A|BC}$ \cite{43}. In \cite{44}, the authors showed that EOF is a monotonic function satisfying $E^2(C^2_{A|B_1B_2\cdots B_{N-1}})\geqslant E^2(\sum_{i=1}^{N-1}C^2_{AB_i})$. For $N-$qubit systems, one has \cite{34}
\begin{equation}\label{eof1}
 E^\alpha_{A|B_1B_2\cdots B_{N-1}}\geqslant E^\alpha_{AB_1}+E^\alpha_{AB_2}+\cdots+E^\alpha_{AB_{N-1}}
\end{equation}
for $\alpha\geqslant\sqrt{2}$, where $E_{A|B_1B_2\cdots B_{N-1}}$ is the entanglement of formation of $\rho$ in bipartite partition $A|B_1B_2\cdots B_{N-1}$, and $E_{AB_i}$, $i=1,2,\cdots,N-1$, is the EOF of the mixed states $\rho_{AB_i}=\mathrm{Tr}_{B_1B_2\cdots B_{i-1},B_{i+1}\cdots B_{N-1}}(\rho)$. It is further proved that for $\alpha\geqslant\sqrt{2}$, if $C_{AB_{i}}\geqslant C_{A|B_{i+1}\cdots B_{N-1}}$ for $i=1,2,\cdots,m$ and $C_{AB_{j}}\leqslant C_{A|B_{j+1}\cdots B_{N-1}}$ for $j= m+1,\cdots,N-2$, $\forall 1 \leqslant m \leqslant N-3$, $N \geqslant 4$, then \cite{35}
\begin{eqnarray}\label{eof2}
&&E^\alpha_{A|B_1B_2\cdots B_{N-1}}\nonumber\\
&&~~~~~~\geqslant E^\alpha_{AB_1}+({\alpha}/{\sqrt{2}}) E^\alpha_{AB_2}\cdots+({\alpha}/{\sqrt{2}})^{m-1}E^\alpha_{AB_m}\nonumber\\
&&~~~~~~~~~+({\alpha}/{\sqrt{2}})^{m+1}(E^\alpha_{AB_{m+1}}+\cdots+E^\alpha_{AB_{N-2}})\nonumber\\
&&~~~~~~~~~+({\alpha}/{\sqrt{2}})^{m}E^\alpha_{AB_{N-1}},
\end{eqnarray}
In fact, generally we can prove the following results.

\textit{Theorem 3}. For any $N$-qubit mixed state $\rho\in \mathds{H}_A\otimes \mathds{H}_{B_1}\otimes\cdots\otimes \mathds{H}_{{B_{N-1}}}$, if
${C_{AB_i}}\geqslant {C_{A|B_{i+1}\cdots B_{N-1}}}$ for $i=1, 2, \cdots, m$, and
${C_{AB_j}}\leqslant {C_{A|B_{j+1}\cdots B_{N-1}}}$ for $j=m+1,\cdots,N-2$, $\forall$ $1\leqslant m\leqslant N-3$, $N\geqslant 4$, the entanglement of formation $E(\rho)$ satisfies
\begin{eqnarray}\label{th3}
&&~~~E^\alpha_{A|B_1B_2\cdots B_{N-1}}\nonumber\\
&&~~~~~~\geqslant E^\alpha_{AB_1}+(2^{t}-1) E^\alpha_{AB_2}\cdots+(2^{t}-1)^{m-1}E^\alpha_{AB_m}\nonumber\\
&&~~~~~~~~~+(2^{t}-1)^{m+1}(E^\alpha_{AB_{m+1}}+\cdots+E^\alpha_{AB_{N-2}})\nonumber\\
&&~~~~~~~~~+(2^{t}-1)^{m}E^\alpha_{AB_{N-1}},
\end{eqnarray}
for $\alpha\geqslant\sqrt{2}$, where $t={\alpha}/{\sqrt{2}}$.

\textit{Proof}. For $\alpha\geqslant\sqrt{2}$, we have
\begin{eqnarray}\label{FA}
 f^{{\alpha}}(x^2+y^2)&&=\left(f^{\sqrt{2}}(x^2+y^2)\right)^t \nonumber\\
 &&\geqslant  \left(f^{\sqrt{2}}(x^2)+f^{\sqrt{2}}(y^2)\right)^t \nonumber\\
 &&\geqslant  \left(f^{\sqrt{2}}(x^2)\right)^t+(2^{t}-1)\left(f^{\sqrt{2}}(y^2)\right)^t\nonumber\\
 && = f^{\alpha}(x^2)+(2^{t}-1)f^{\alpha}(y^2),
\end{eqnarray}
where the first inequality is due to the inequality (\ref{F2}), and the second inequality is obtained from a similar consideration in the proof of the second inequality in (\ref{le2}).

Let $\rho=\sum\limits_ip_i|\psi_i\rangle\langle\psi_i|\in \mathds{H}_A\otimes \mathds{H}_{B_1}\otimes\cdots\otimes \mathds{H}_{{B_N-1}}$ be the optimal decomposition of $E_{A|B_1B_2\cdots B_{N-1}}(\rho)$ for the $N$-qubit mixed state $\rho$; then we have
\begin{eqnarray*}\label{ED}
E_{A|B_1B_2\cdots B_{N-1}}(\rho)&&=\sum_ip_iE_{A|B_1B_2\cdots B_{N-1}}(|\psi_i\rangle)\nonumber\\
&&=\sum_ip_if\left(C^2_{A|B_1B_2\cdots B_{N-1}}(|\psi_i\rangle)\right)\nonumber\\
&&\geqslant f\left(\sum_ip_iC^2_{A|B_1B_2\cdots B_{N-1}}(|\psi_i\rangle)\right)\nonumber\\
&&\geqslant f\left(\left[\sum_ip_iC_{A|B_1B_2\cdots B_{N-1}}(|\psi_i\rangle)\right]^2\right)\nonumber\\
&& \geqslant f\left(C^2_{A|B_1B_2\cdots B_{N-1}}(\rho)\right),\nonumber
\end{eqnarray*}
where the first inequality is due to the fact that $f(x)$ is a convex function. The second inequality is due to the Cauchy-Schwarz inequality: $(\sum\limits_ix_i^2)^{\frac{1}{2}}(\sum\limits_iy_i^2)^{\frac{1}{2}}\geqslant\sum\limits_ix_iy_i$, with $x_i=\sqrt{p_i}$ and $y_i=\sqrt{p_i}C_{A|B_1B_2\cdots B_{N-1}}(|\psi_i\rangle)$. Due to the definition of concurrence and that $f(x)$ is a monotonically increasing function, we obtain the third inequality. Therefore, we have
\small
\begin{flalign*}
&E^\alpha_{A|B_1B_2\cdots B_{N-1}}(\rho)\nonumber\\
&~~~\geqslant f^\alpha(C^2_{AB_1}+C^2_{AB_2}+\cdots+C^2_{AB_{m-1}})\nonumber\\
&~~~\geqslant f^{\alpha}(C^2_{AB_1})+(2^{t}-1) f^{\alpha}(C^2_{AB_2})+\cdots+(2^{t}-1)^{m-1} f^{\alpha}(C^2_{AB_m})\nonumber\\
&~~~~~~~+(2^{t}-1)^{m+1}(f^{\alpha}(C^2_{AB_{m+1}})+\cdots+f^{\alpha}(C^2_{AB_{N-2}}))\nonumber\\
&~~~~~~~+(2^{t}-1)^{m}f^{\alpha}(C^2_{AB_{N-1}})\nonumber\\
&~~~=E^\alpha_{A|B_1}+(2^{t}-1) E^\alpha_{AB_2}+\cdots+(2^{t}-1)^{m-1}E^\alpha_{AB_m}\nonumber\\
&~~~~~~~+(2^{t}-1)^{m+1}(E^\alpha_{AB_{m+1}}+\cdots+E^\alpha_{AB_{N-2}})\nonumber\\
&~~~~~~~+(2^{t}-1)^{m}E^\alpha_{AB_{N-1}},\nonumber
\end{flalign*}
\normalsize
where we have used the monogamy inequality in (\ref{mo1}) for $N$-qubit states $\rho$ to obtain the first inequality. By using (\ref{FA}) and the similar consideration in the proof of Theorem 1, we get the second inequality. Since for any $2\otimes2$ quantum state $\rho_{AB_i}$, $E(\rho_{AB_i})=f\left[C^2(\rho_{AB_i})\right]$, one gets the last equality. ~~~~~~~~~~~~~~~~~~~~~~~~~~~~~~~~~~~~~~~~~~~~~~~~~~~~~~~~~~~~~~~~~~~~~~$\blacksquare$

As for $(2^{ {\alpha}/{\sqrt{2}}}-1)\geqslant{\alpha}/{\sqrt{2}}$ for $\alpha\geqslant\sqrt{2}$, (\ref{th3}) is obviously tighter than (\ref{eof1}), (\ref{eof2}). Moreover, similar to the concurrence, for the case that ${C_{AB_i}}\geqslant {C_{A|B_{i+1}\cdots B_{N-1}}}$ for all $i=1, 2, \cdots, N-2$, we have a simple tighter monogamy relation for the entanglement of formation:

\textit{Theorem4}.
If ${C_{AB_i}}\geqslant {C_{A|B_{i+1}\cdots B_{N-1}}}$ for all $i=1, 2, \cdots, N-2$, we have
\begin{eqnarray}\label{th4}
E^\alpha_{A|B_1B_2\cdots B_{N-1}}&&\geqslant E^\alpha_{AB_1}+(2^{{\alpha}/{\sqrt{2}}}-1) E^\alpha_{AB_2}+\cdots\nonumber\\
&&~~~~+(2^{{\alpha}/{\sqrt{2}}}-1)^{N-2}E^\alpha_{AB_{N-1}}
\end{eqnarray}
for $\alpha\geqslant\sqrt{2}$.

{\it Example 2}. Let us consider the $W$ state, $|W\rangle=\frac{1}{\sqrt{3}}(|100\rangle+|010\rangle+|001\rangle).$ We have $E_{AB}=E_{AC}=0.550048$, $E_{A|BC}=0.918296$, and then $E_{A|BC}^{\alpha}=(0.918296)^{\alpha}$, $E^{\alpha}_{AB}+E^{\alpha}_{AC}=2(0.550048)^{\alpha}$, $E^{\alpha}_{AB}+\frac{\alpha}{\sqrt{2}}E^{\alpha}_{AC}=(1+\frac{\alpha}{\sqrt{2}})(0.550048)^{\alpha}$, $E^{\alpha}_{AB}+(2^{\frac{\alpha}{\sqrt{2}}}-1)E^{\alpha}_{AC}=2^{\frac{\alpha}{\sqrt{2}}}(0.550048)^{\alpha}$. It is easily verified that our results are better than the results in \cite{34} and \cite{35} for $\alpha\geqslant\sqrt{2}$; see Fig 2.
\begin{figure}
\centering
    \includegraphics[width=7cm]{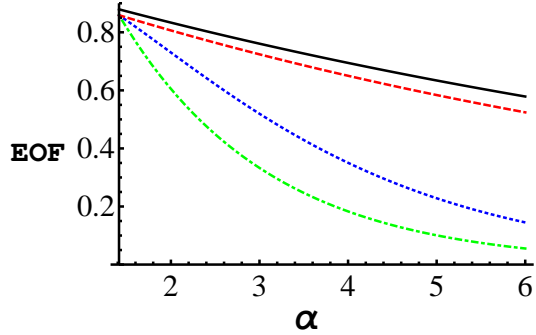}\\
  \caption{Behavior of the EOF of $|W\rangle$ and its lower bound, which are functions of $\alpha$ plotted. The black solid line represents the EOF of the state $|W\rangle$ in Example 2, the red dashed line represents the lower bound from our result, and the blue dotted (green dot-dashed) line represents the lower bound from the result in \cite{35} (\cite{34}).}\label{2}
\end{figure}

\section{TIGHTER MONOGAMY RELATIONS FOR NEGATIVITY }

Another well-known quantifier of bipartite entanglement is the negativity. Given a bipartite state $\rho_{AB}$ in $\mathds{H}_A\otimes \mathds{H}_B$, the negativity is defined by \cite{45} $N(\rho_{AB})=(||\rho_{AB}^{T_A}||-1)/2$, where $\rho_{AB}^{T_A}$ is the partial transpose with respect to the subsystem $A$, and $||X||$ denotes the trace norm of $X$, i.e $||X||=\mathrm{Tr}\sqrt{XX^\dag}$. Negativity is a computable measure of entanglement and is a convex function of $\rho_{AB}$. It vanishes if and only if $\rho_{AB}$ is separable for the $2\otimes2$ and $2\otimes3$ systems \cite{46}. For the purpose of discussion, we use the following definition of negativity, $ N(\rho_{AB})=||\rho_{AB}^{T_A}||-1$.
For any bipartite pure state $|\psi\rangle_{AB}$, the negativity $ N(\rho_{AB})$ is given by
$N(|\psi\rangle_{AB})=2\sum\limits_{i<j}\sqrt{\lambda_i\lambda_j}=(\mathrm{Tr}\sqrt{\rho_A})^2-1$,
where $\lambda_i$ are the eigenvalues for the reduced density matrix of $|\psi\rangle_{AB}$. For a mixed state $\rho_{AB}$, the convex-roof extended negativity (CREN) is defined as
\begin{equation}\label{nc}
 N_c(\rho_{AB})=\mathrm{min}\sum_ip_iN(|\psi_i\rangle_{AB}),
\end{equation}
where the minimum is taken over all possible pure-state decompositions $\{p_i,~|\psi_i\rangle_{AB}\}$ of $\rho_{AB}$. CREN gives a perfect discrimination of positive partial transposed bound entangled states and separable states in any bipartite quantum system \cite{47,48}.

Let us consider the relation between CREN and concurrence. For any bipartite pure state    $|\psi\rangle_{AB}$ in a $d\otimes d$ quantum system with Schmidt rank 2,
$|\psi\rangle_{AB}=\sqrt{\lambda_0}|00\rangle+\sqrt{\lambda_1}|11\rangle$,
one has
$N(|\psi\rangle_{AB})=\parallel|\psi\rangle\langle\psi|^{T_B}\parallel-1=2\sqrt{\lambda_0\lambda_1}
=\sqrt{2(1-\mathrm{Tr}\rho_A^2)}=C(|\psi\rangle_{AB})$. In other words, negativity is equivalent to concurrence for any pure state with Schmidt rank 2, and consequently it follows that for any two-qubit mixed state $\rho_{AB}=\sum p_i|\psi_i\rangle_{AB}\langle\psi_i|$,
\begin{eqnarray}\label{N1}
 N_c(\rho_{AB})&&=\mathrm{min}\sum_ip_iN(|\psi_i\rangle_{AB})\\ \nonumber
&&=\mathrm{min}\sum_ip_iC(|\psi_i\rangle_{AB})\\ \nonumber
&&=C(\rho_{AB}).
\end{eqnarray}

With a similar consideration of concurrence, we obtain the following result.

\textit{Theorem 5}. For any $N$-qubit state $\rho\in \mathds{H}_A\otimes \mathds{H}_{B_1}\otimes\cdots\otimes \mathds{H}_{B_{N-1}}$, if
${N_c}_{AB_i}\geqslant {N_c}_{A|B_{i+1}\cdots B_{N-1}}$ for $i=1, 2, \cdots, m$, and
${N_c}_{AB_j}\leqslant {N_c}_{A|B_{j+1}\cdots B_{N-1}}$ for $j=m+1,\cdots,N-2$,
$\forall$ $1\leqslant m \leqslant N-3$, $N\geqslant 4$, we have
\begin{eqnarray}\label{la6}
&&~~~~~~{N^\alpha_c}_{A|B_1B_2\cdots B_{N-1}} \nonumber\\
&&~~~~~~\geqslant {N^\alpha_c}_{AB_1}+(2^{\frac{\alpha}{2}}-1) {N^\alpha_c}_{AB_2}+\cdots+(2^{\frac{\alpha}{2}}-1)^{m-1}{N^\alpha_c}_{AB_m}\nonumber\\
 &&~~~~~~~~~+(2^{\frac{\alpha}{2}}-1)^{m+1}({N^\alpha_c}_{AB_{m+1}}
 +\cdots+{N^\alpha_c}_{AB_{N-2}})\nonumber\\
 &&~~~~~~~~~+(2^{\frac{\alpha}{2}}-1)^{m}{N^\alpha_c}_{AB_{N-1}}
\end{eqnarray}
for all $\alpha\geqslant2$.

In Theorem 5 we have assumed that
some ${N_c}_{AB_i}\geqslant {N_c}_{A|B_{i+1}\cdots B_{N-1}}$ and some
${N_c}_{AB_j}\leq {N_c}_{A|B_{j+1}\cdots B_{N-1}}$ for the $2\otimes2\otimes\cdots\otimes2$ mixed state $\rho\in \mathds{H}_A\otimes \mathds{H}_{B_1}\otimes\cdots\otimes \mathds{H}_{{B_{N-1}}}$.
If all ${N_c}_{AB_i}\geqslant {N_c}_{A|B_{i+1}\cdots B_{N-1}}$ for $i=1, 2, \cdots, N-2$, then we have
the following conclusion:

\textit{Theorem 6}.
If ${N_c}_{AB_i}\geqslant {N_c}_{A|B_{i+1}\cdots B_{N-1}}$ for all $i=1, 2, \cdots, N-2$, then we have
\begin{eqnarray}\label{}
{N_c}^\alpha_{A|B_1\cdots B_{N-1}}&&\geqslant {N_c}^\alpha_{AB_1}+(2^{\frac{\alpha}{2}}-1) {N_c}^\alpha_{AB_2}+\cdots\nonumber\\
&&~~~~+(2^{\frac{\alpha}{2}}-1)^{N-2}{N_c}^\alpha_{AB_{N-1}}.
\end{eqnarray}

\textit{Example 3}. Let us consider again the three-qubit state $|\psi\rangle$ $(9)$. From the definition of CREN, we have ${N_c}_{A|BC} = 2 \lambda_{0}\sqrt{\lambda_{2}^{2}+\lambda_{3}^{2}+\lambda_{4}^{2}}$, ${N_c}_{AB}=2\lambda_0 \lambda_2$, and ${N_c}_{AC}=2\lambda_0 \lambda_3$. Set $\lambda_0=\lambda_1=\lambda_2=\lambda_3=\lambda_4=\frac{\sqrt{5}}{5}$. One gets ${N_c}^\alpha_{A|BC}=(\frac{2\sqrt{3}}{5})^{\alpha}$, ${N_c}^\alpha_{AB}+{N_c}^\alpha_{AC}=2(\frac{2}{5})^{\alpha}$, ${N_c}^\alpha_{AB}+\frac{\alpha}{2}{N_c}^\alpha_{AC}=(1+\frac{\alpha}{2})(\frac{2}{5})^{\alpha}$, ${N_c}^\alpha_{AB}+(2^{\frac{\alpha}{2}}-1){N_c}^\alpha_{AC}=2^{\frac{\alpha}{2}}(\frac{2}{5})^{\alpha}$. One can see that our result is better than the results in $\cite{34}$ and $\cite{36}$ for $\alpha\geqslant2$; see Fig. 3.

\begin{figure}
\centering
    \includegraphics[width=7cm]{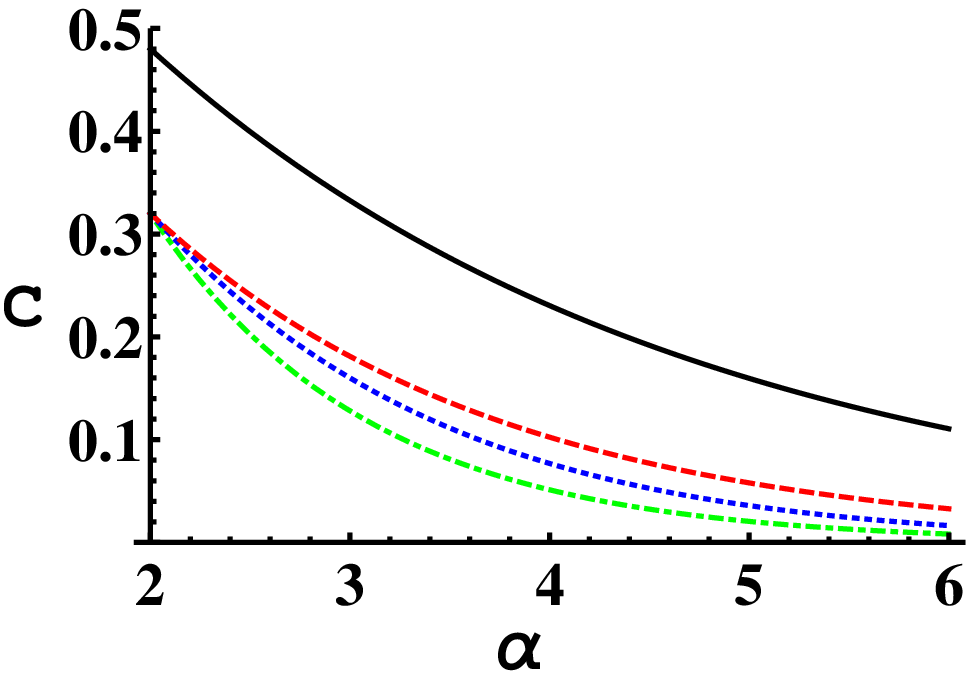}\\
  \caption{Behavior of the concurrence of $|\psi\rangle$ and its lower bound, which are functions of $\alpha$ plotted. The black solid line represents the concurrence of $|\psi\rangle$ in Example 3, the red dashed line represents the lower bound from our result, and the blue dotted (green dot-dashed) line represents the lower bound from the result in \cite{36} (\cite{34}).}
\end{figure}

\section{Tighter monogamy relations for Tsallis-q entanglement}

For a bipartite pure state $|\psi\rangle_{AB}$, the Tsallis-$q$ entanglement is defined by $\cite{24}$
\begin{eqnarray}\label{tq}
T_q(|\psi\rangle_{AB})=S_q(\rho_A)=\frac{1}{q-1}(1-\mathrm{tr}\rho_A^q),
\end{eqnarray}
for any $q > 0$ and $q \ne 1$. If $q$ tends to 1, $T_q(\rho)$ converges to the von Neumann entropy, $\lim_{q\to1} T_q(\rho)=-\mathrm{tr}\rho\ln\rho=S_q(\rho)$. For a bipartite mixed state $\rho_{AB}$, Tsallis-$q$ entanglement is defined via the convex-roof extension,
$
T_q(\rho_{AB})=\min\sum_ip_iT_q(|\psi_i\rangle_{AB}),
$
with the minimum taken over all possible pure-state decompositions of $\rho_{AB}$.

In \cite{49}, the author has proved an analytic relationship between Tsallis-$q$ entanglement and concurrence for $\frac{5-\sqrt{13}}{2}\leq q\leq \frac{5+\sqrt{13}}{2}$,
\begin{eqnarray}\label{an1}
T_q(|\psi\rangle_{AB})=\textsl{g}_q(C^2(|\psi\rangle_{AB})),
\end{eqnarray}
where the function $\textsl{g}_q(x)$ is defined as
\small
\begin{eqnarray}\label{an2}
\textsl{g}_q(x)=\frac{1}{q-1}\left[1-\left(\frac{1+\sqrt{1-x}}{2}\right)^q-\left(\frac{1-\sqrt{1-x}}{2}\right)^q\right].
\end{eqnarray}
\normalsize
It has been shown that $T_q(|\psi\rangle)=\textsl{g}_q\left(C^2(|\psi\rangle)\right)$ for $2\otimes m~(m\geqslant2)$ pure state $|\psi\rangle$, and  $T_q(\rho)=\textsl{g}_q\left(C^2(\rho)\right)$ for two-qubit mixed state $\rho$ in \cite{24}. Hence, (\ref{an1}) holds for any $q$ such that $\textsl{g}_q(x)$ in (\ref{an2}) is monotonically increasing and convex. In particular, $\textsl{g}_q(x)$ satisfies the following relations for $2\leqslant q\leqslant 3$:
\begin{eqnarray}\label{gx}
\textsl{g}_q(x^2+y^2)\geqslant \textsl{g}_q(x^2)+\textsl{g}_q^2(y^2).
\end{eqnarray}

The Tsallis-$q$ entanglement satisfies $\cite{24}$
\begin{eqnarray}\label{stq1}
{T_q}_{A|B_1B_2\cdots B_{N-1}}\geqslant\sum_{i=1}^{N-1}{T_q}_{AB_i},
\end{eqnarray}
where $i=1,2,\cdots N-1$, $2\leqslant q\leqslant 3$. It is further proved in \cite{49}
\begin{eqnarray}\label{stq2}
{T_q^2}_{A|B_1B_2\cdots B_{N-1}}\geqslant\sum_{i=1}^{N-1}{T_q^2}_{AB_i},
\end{eqnarray}
with $\frac{5-\sqrt{13}}{2}\leqslant q\leqslant \frac{5+\sqrt{13}}{2}$.
 In fact, generally we can prove the following results.

\textit{Theorem 7}. For an arbitrary $N$-qubit mixed state $\rho_{AB_1\cdots B_{N-1}}$, if
${C_{AB_i}}\geqslant {C_{A|B_{i+1}\cdots B_{N-1}}}$ for $i=1, 2, \cdots, m$, and
${C_{AB_j}}\leqslant {C_{A|B_{j+1}\cdots B_{N-1}}}$ for $j=m+1,\cdots,N-2$,
$\forall$ $1\leqslant m\leqslant N-3$, $N\geqslant 4$, then the $\alpha$th power of Tsallis-$q$ satisfies the monogamy relation
\begin{eqnarray}\label{th6}
&&{T_q}^\alpha_{A|B_1B_2\cdots B_{N-1}}\nonumber\\
&&~~~\geqslant{T_q}^\alpha_{AB_1}+(2^{{\alpha}}-1) {T_q}^\alpha_{AB_2}+\cdots+(2^{{\alpha}}-1)^{m-1}{T_q}^\alpha_{AB_m}\nonumber\\
&&~~~~~~+(2^{{\alpha}}-1)^{m+1}({T_q}^\alpha_{AB_{m+1}}+\cdots+{T_q}^\alpha_{AB_{N-2}})\nonumber\\
&&~~~~~~+(2^{{\alpha}}-1)^{m}{T_q}^\alpha_{AB_{N-1}},
\end{eqnarray}
\normalsize
where $\alpha\geqslant 1$, ${T_q}_{A|B_1B_2\cdots B_{N-1}}$ quantifies the Tsallis-$q$ entanglement in the partition $A|B_1B_2\cdots B_{N-1}$ and ${T_q}_{AB_i}$ quantifies that in two-qubit subsystem $AB_i$ with $2\leqslant q\leqslant 3$.

\textit{Proof}. For $\alpha\geqslant1$, we have
\begin{eqnarray}\label{pf61}
 \textsl{g}_q^{{\alpha}}(x^2+y^2)
 &&\geqslant \left(\textsl{g}_q(x^2)+\textsl{g}_q(y^2)\right)^{\alpha} \nonumber \\
 &&\geqslant \textsl{g}_q^{\alpha}(x^2)+(2^{\alpha}-1)\textsl{g}_q^{\alpha}(y^2),
\end{eqnarray}
where the first inequality is due to the inequality (\ref{gx}), and the second inequality is obtained from a similar consideration in the proof of the second inequality in (\ref{le2}).

Let $\rho=\sum\limits_ip_i|\psi_i\rangle\langle\psi_i|\in \mathds{H}_A\otimes \mathds{H}_{B_1}\otimes\cdots\otimes \mathds{H}_{{B_N-1}}$ be the optimal decomposition for the $N$-qubit mixed state $\rho$; then we have
\begin{eqnarray}\label{pf62}
&&{T_q}_{A|B_1B_2\cdots B_{N-1}}(\rho)\nonumber\\
&&=\sum_ip_iT_q(|\psi_i\rangle_{A|B_1B_2\cdots B_{N-1}})\nonumber\\
&&=\sum_ip_i\textsl{g}_q\left[C^2_{A|B_1B_2\cdots B_{N-1}}(|\psi_i\rangle)\right]\nonumber\\
&&\geqslant \textsl{g}_q\left[\sum_ip_iC^2_{A|B_1B_2\cdots B_{N-1}}(|\psi_i\rangle)\right]\nonumber\\
&&\geqslant \textsl{g}_q\left(\left[\sum_ip_iC_{A|B_1B_2\cdots B_{N-1}}(|\psi_i\rangle)\right]^2\right)\nonumber\\
&&=\textsl{g}_q\left[C^2_{A|B_1B_2\cdots B_{N-1}}(\rho)\right],\nonumber\\
\end{eqnarray}
where the first inequality is due to the fact that $\textsl{g}_q(x)$ is a convex function. The second inequality is due to the Cauchy-Schwarz inequality: $(\sum\limits_ix_i^2)^{\frac{1}{2}}(\sum\limits_iy_i^2)^{\frac{1}{2}}\geqslant\sum\limits_ix_iy_i$, with $x_i=\sqrt{p_i}$ and $y_i=\sqrt{p_i}C_{A|B_1B_2\cdots B_{N-1}}(|\psi_i\rangle)$. Due to the definition of Tsallis-$q$ entanglement and that $\textsl{g}_q(x)$ is a monotonically increasing function, we obtain the third inequality.
Therefore, we have
\small
\begin{eqnarray}\label{pf63}
&&{T_q^\alpha}_{A|B_1B_2\cdots B_{N-1}}(\rho)\nonumber\\
&&~~~\geqslant \textsl{g}_q^\alpha\left[\sum_iC^2(\rho_{AB_i})\right]\nonumber\\
&&~~~\geqslant {\textsl{g}_q}^\alpha(C_{AB_1})+(2^{\alpha}-1) {\textsl{g}_q}^\alpha(C_{AB_2})+\cdots\nonumber\\
&&~~~~~~~+(2^{\alpha}-1)^{m-1}{\textsl{g}_q}^\alpha(C_{AB_m})\nonumber\\
&&~~~~~~~+(2^{\alpha}-1)^{m+1}\left({\textsl{g}_q}^\alpha(C_{AB_{m+1}})+\cdots+{\textsl{g}_q}^\alpha(C_{AB_{N-2}})\right)\nonumber\\
&&~~~~~~~+(2^{\alpha}-1)^{m}{\textsl{g}_q}^\alpha(C_{AB_{N-1}})\nonumber\\
&&~~~= {T_q}^\alpha_{AB_1}+(2^{\alpha}-1) {T_q}^\alpha_{AB_2}+\cdots+(2^{\alpha}-1)^{m-1}{T_q}^\alpha_{AB_m}\nonumber\\
&&~~~~~~~+(2^{\alpha}-1)^{m+1}({T_q}^\alpha_{AB_{m+1}}+\cdots+{T_q}^\alpha_{AB_{N-2}})\nonumber\\
&&~~~~~~~+(2^{\alpha}-1)^{m}{T_q}^\alpha_{AB_{N-1}},
\end{eqnarray}
\normalsize
where we have used the monogamy inequality in (\ref{mo1}) for $N$-qubit states $\rho$ to obtain the first
inequality. By using (\ref{pf61}) and the similar consideration in the proof of Theorem 1, we get the second
inequality. Since for any $2\otimes2$ quantum state $\rho_{AB_i}$, $T_q(\rho_{AB_i})=\textsl{g}_q\left[C^2(\rho_{AB_i})\right]$, one gets the last equality. ~~~~~~~~~~~~~~~~~~~~~~~~~~~~~~~~~~~~~~~~~~~~~~~~~~~~~~~~~~~~~~~~~~~~~~~~$\blacksquare$

\textit{Example 4}. Let us consider again the three-qubit state $|\psi\rangle$ $(9)$. From the definition of Tsallis-$q$ entanglement, when $q=2$, we have $T_{2A|BC} = 2 \lambda_{0}^2(\lambda_{2}^{2}+\lambda_{3}^{2}+\lambda_{4}^{2})$, $T_{2AB}=2\lambda_0^2 \lambda_2^2$, and $T_{2AC}=2\lambda_0^2 \lambda_3^2$. Set $\lambda_0=\lambda_1=\lambda_2=\lambda_3=\lambda_4=\frac{\sqrt{5}}{5}$. One gets ${T_2}^\alpha_{A|BC}=(\frac{6}{25})^{\alpha}$, ${T_2}^\alpha_{AB}+{T_2}^\alpha_{AC}=2(\frac{2}{25})^{\alpha}$, ${T_2}^\alpha_{AB}+(2^{\frac{\alpha}{2}}-1){T_2}^\alpha_{AC}=2^{\alpha}(\frac{2}{25})^{\alpha}$. One can see that our result is better than that in $\cite{34}$ for $\alpha\geqslant2$; see Fig. 4.

\begin{figure}
\centering
    \includegraphics[width=7cm]{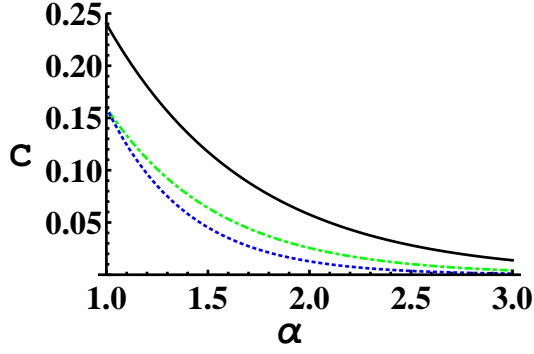}\\
  \caption{Behavior of the concurrence of $|\psi\rangle$ and its lower bound, which are functions of $\alpha$ plotted. The black solid line represents the concurrence of $|\psi\rangle$ in Example 4, the green dot-dashed line represents the lower bound from our result, and the blue dotted line represents the lower bound from the result in  \cite{24} .}
\end{figure}

\section{conclusion}

Entanglement monogamy is a fundamental property of multipartite entangled states. We have presented monogamy relations related to the $\alpha$ power of concurrence $C$, entanglement of formation $E$, negativity $N_c$, and Tsallis-$q$ entanglement $T_q$, which are tighter, at least for some classes of quantum states, than the existing entanglement monogamy relations for $\alpha > 2$, $\alpha > \sqrt{2}$, $\alpha > 2$, $\alpha > 1$, respectively. The necessary conditions that our inequalities are strictly tighter can been seen from our monogamy relations. For instance, (8) s tighter than the existing ones for $\alpha > 2$, for all quantum states where at least one of the $C_{AB_i}$'s ($i=2,\cdots,N-1$) is not zero, which excludes the fully separable states that have no entanglement distribution at all among the subsystems. Another case that $C_{AB_i}=0$ for all $i=2,\cdots,N-1$ is the $N$-qubit GHZ state $\cite{50}$, which is genuine multipartite entangled. However, for the genuine entangled $N$-qubit $W$ state $\cite{51}$, one has $C_{AB_i}=\frac{2}{N}$, $i=2,\cdots,N-1$, In general, most of states have at least one nonzero $C_{AB_i}$ ($i=2,\cdots,N-1$).

Monogamy relations characterize the distributions of entanglement in multipartite systems. Tighter monogamy relations imply finer characterizations of the entanglement distribution. Our approach may also be used to further study the monogamy properties related to other quantum correlations.

\bigskip
\noindent{\bf Acknowledgments}\, \, This work is supported by the NSF of China under Grant No. 11675113 and  is supported by the Research Foundation for Youth Scholars of Beijing Technology and Business University QNJJ2017-03.

\end{document}